\documentclass[twocolumn,showpacs,preprintnumbers,amsmath,amssymb]{revtex4}

\usepackage{color}
\usepackage{graphicx}
\usepackage{dcolumn}
\usepackage{bbm}
\usepackage{bm}
\usepackage{overpic}

\begin{document}

\title{AC-driven Quantum Phase Transition in the Lipkin-Meshkov-Glick Model}
\author{G. Engelhardt}
\email{georgt@itp.tu-berlin.de}

\author{V. M. Bastidas}
\email{victor@physik.tu-berlin.de}
\author{C. Emary}
\author{T. Brandes}

\affiliation{%
Institut f\"ur Theoretische Physik, Technische Universit\"at Berlin, Hardenbergstr. 36, 10623 Berlin, Germany}%

\begin{abstract}
We establish a set of nonequilibrium quantum phase transitions in the Lipkin-Meshkov-Glick model  under monochromatic modulation of the inter-particle interaction.
We show that the external driving induces a rich phase diagram that characterizes the multistability in the system. Interestingly, the number of stable configurations can be tuned by increasing the amplitude of the driving field.
Furthermore, by studying the quantum evolution, we demonstrate that
the system exhibits a set of quantum phases that correspond to dynamically stabilized states.
\end{abstract}


\maketitle
A quantum phase transition (QPT) is a drastic change of state of a many-body system which occurs at zero temperature \cite{Sachdev}. In contrast to thermodynamic phase transitions---which are driven by thermal fluctuations, QPTs are entirely driven by quantum fluctuations. 
Rather recently, there is an increasing interest in the study of many-particle systems
under the an external variation of the parameters \cite{Keitel,Altland,Polkovnikov,Zurek,Hangui,Goren1,Holthaus,Goren2,Arimondo}.
Furthermore, recent works explore the quantum dynamics of topological states of matter  
controlled by a driving field \cite{AokiTanaka, Lindner,Tanaka, Zoller, Reynoso}.
Recent investigations show that by means of monochromatic fields, unconventional quantum phases in the Dicke model \cite{Bastidas3} and the quantum Ising model \cite{Bastidas4} can be generated, which are unaccessible in their undriven counterparts.

In contrast to undriven QPTs, transitions to excited levels are induced as a result of the external driving, even if the system is prepared initially in the ground state of the undriven model. Therefore, signatures of quantum criticality can also appear in excited states.
A similar issue has been explored in undriven models, where a recent line of research are excited-state phase transitions, e.g. in mean-field type QPTs such as the Lipkin-Meshkov-Glick \cite{RibeiroESQPT} or the Dicke superradiance \cite{Relanho} model. Furthermore, the connection between chaos and criticality \cite{Brandes} has been studied in detail for the interacting boson model \cite{Leviatan}.

Despite of the original motivation of the Lipkin-Meshkov-Glick (LMG) model as a toy model to test approximation methods in manybody physics \cite{LipkinMeshkovGlick}, currently it constitutes an active field of research and a natural scenario to study the relation between QPTs and spin squeezing \cite{NoriSpinSqueez} and quantum Fisher information as a resource for high-precision quantum estimation \cite{Wang}. Furthermore, rather recently, experimental realizations of the LMG model in optical cavity QED \cite{Parkins1,Parkins2} and circuit QED \cite{Larson} have been suggested.

In this paper we study a  driven version of the LMG model in which we assume a time-dependent inter-particle interaction.
In comparison with previous works which considered dynamical properties of the LMG model under an adiabatic change of parameters of the system across the quantum critical point \cite{Mosseri}, the effect  of fast and slow quenches of the transverse field \cite{SantoroLMG}, and a periodically-driven uniaxial LMG model \cite{Hangui,Chakrabarti,Dykman}, we address here the fundamental issue of driving-induced QPTs in the LMG model.

The current experimental feasibilities allow us to control externally the parameters of many-body systems. For example, the kicked top, has been realized experimentally in an ensemble of laser-cooled Cs atoms \cite{Chaudhury}. Furthermore, an optical realization of the uniaxial driven LMG in photonic lattices \cite{Longhi} and by using superconducting charge qubits connected in parallel to a common superconductor inductance \cite{Liang} have been suggested.

We show that when the driving is near resonance with the excitation energy of the undriven system the nature of criticality changes dramatically. Additionally, we find that the external driving induces stable configurations which do not have analogue in the undriven system. This gives rise to a novel route of experimental studies exploring  the characteristics of criticality under nonequilibrium conditions.

A more specific outline of our paper is as follows. In section \ref{SectionI} we describe the fundamentals on the QPT in the undriven LMG model and construct a bosonized effective Hamiltonian which allows us to describe the stability properties of the symmetric phase. In section \ref{SectionII} we find an effective Hamiltonian by using the rotating wave approximation (RWA) and describe the quantum evolution of the observables to understand
the characteristics of the novel nonequilibrium metastable phases. Finally, conclusive remarks are given in section \ref{SectionIII}.

%%%%%%%%
\section{Quantum resonances in the LMG  model. \label{SectionI}}
%%%%%%%%
The periodically-driven Lipkin-Meshkov-Glick model describes the dynamics of $N$ interacting two-level systems in a transverse local field 
%%%
\begin{equation}
      \label{LMGDriven}
            \hat{H}(t) = -hJ_z-\frac{1}{N}\left(\gamma^{x}(t) {J_x}^2+\gamma^y{J_y}^2\right)
      ,
\end{equation}
%%%
where $J_{\alpha}=\frac{1}{2} \sum_{i=1} ^{N} \sigma_{\alpha} ^{(i)} $ denote collective angular momentum operators  and $\sigma_{\alpha} ^{(i)} $ are the Pauli matrices with $\alpha \in \{x,y,z\}$. These operators satisfy the $SU\left(2\right)$ algebra $\left[ J_{\alpha},J_{\beta} \right] = i \epsilon_{\alpha \beta \gamma } J_{\gamma} $. In the following we shall consider $h<0$, and a monochromatic modulation of the inter-particle interaction with a static contribution: 
$\gamma^x(t)= \gamma^x_{0} + \gamma^x_{1} \cos \Omega t$. \\
In this paper we consider Dicke states i.e., the  whole analysis is reduced to the Hilbert subspace characterized by a maximal total angular momentum  $j=N/2$. The Hamiltonian Eq. \eqref{LMGDriven} has a conserved parity
%%%
\begin{equation}
      \label{Parity}
            \hat{\Pi}=\exp(i\pi(J_z+j))
      ,
\end{equation}
such that $[\hat{H}(t),\hat{\Pi}]=0$.
Our aim in this paper is to study the new aspects of criticality under the effect of driving.
In this section we provide the basics on the formalism used to describe the effective Hamiltonian for the symmetric phase of the LMG  model. In particular, we find a resonance condition related to $m$-photon processes under the effect of driving.
% % % % % % % %
\subsection{The QPT in the undriven LMG model}
% % % % % % % % %
In the case of the undriven LMG model $(\gamma^{x}_{1}=0)$, an analytic study of the energy surface in the thermodynamic limit $N\rightarrow \infty$ leads to a phase diagram in the $(\gamma^{x}_{0},\gamma^{y})$-plane, which is divided into four regions depending on the geometry of the surface. These zones are distinguished from each other by the number of minima, maxima and saddle points, which are related to non-analyticities of the density of states \cite{RibeiroESQPT}. As a consequence of the symmetry of the LMG Hamiltonian, it is sufficient to consider the region with $\left|\gamma^{x}_{0}\right| < \gamma^{y}$. In the region $\left|\gamma^{x}_{0}\right|<\gamma^{y}<-h$ the  energy landscape has a single global minimum, whereas in the regions with $\left|\gamma^{x}_{0}\right|<-h<\gamma^{y}$ the surface has two global minima. By crossing the critical line $\gamma^{y} = -h$, the single global minimum splits in 
two global minima and the system exhibits a continuous transition from a symmetric state to a symmetry-broken state, i.e., a second-order QPT.
%
%%%%%%%%%%%%
\subsection{Effective bosonized Hamiltonian for the symmetric phase}
%%%%%%%%%%%%
We begin our analysis by investigating the stability of the symmetric phase under the effect of an external driving.
To this end, we construct a symmetric phase effective Hamiltonian in the same way as in Ref.~\cite{Bastidas3}: we make a Holstein-Primakoff representation of the angular momentum algebra in terms of bosonic operators $b,b^\dag$
%%%
\begin{align}
	    J_z&=b^{\dagger}b-\frac{N}{2} ,\label{HPZ} \\ 
	    J_{+}&=b^{\dagger}~\sqrt{N-b^{\dagger}b} ,\label{HPMas} \\ 
            J_{-}&=\sqrt{N-b^{\dagger}b}~~b 
      , \label{HPMen}
\end{align}
%%%
and take the thermodynamic limit $N\rightarrow \infty$, assuming $b/N \rightarrow 0$.
The result is
a bosonized Hamiltonian for the symmetric phase
%%%
\begin{equation}
      \label{BosHamSymmetricPhase}
            \hat{H}_{S}(t) = -h\ b^{\dagger}b-\frac{1}{4}[\gamma^{x}(t) (b^{\dagger}+b)^2-\gamma^{y}(b^{\dagger}-b)^2]+\frac{N h}{2}
      ,
\end{equation}
%%%
which resemble the bosonized Hamiltonian for the undriven LMG model \cite{VidalFSExp, VidalFullConSpin}, in this case, however, the Hamiltonian is characterized by a time dependent squeezing parameter \cite{NoriSpinSqueez}. Previous works have used the Holstein-Primakoff transformation
to study the finite-size exponents in the LMG model \cite{VidalFSExp}, entanglement measurements in fully connected spin models \cite{VidalFullConSpin}, and to investigate collective spin systems at high temperatures \cite{Hamdouni}.
By introducing the coordinate operators in Eq. \eqref{BosHamSymmetricPhase}
%%%
\begin{align}
      \label{Cuadratures}
	    \hat{q} &= \sqrt{-\frac{1}{2h}\left(\frac{h}{h+\gamma^{y}}\right)}(b^{\dagger}+b) ,\\
	    \hat{p} &= i\sqrt{-\frac{h}{2}\left(1+\frac{\gamma^{y}}{h}\right)}(b^{\dagger}-b)
      ,    
\end{align}
%%%
we obtain the Hamiltonian of a parametrically-driven harmonic oscillator \cite{Arnold,KohlerDittrich,Zerbe,ThorwartHaenggi,PeanoDykman}
%%%
\begin{equation}
      \label{BosHamSymmetricPhaseCuadratures}
            \hat{H}_{S}(t) = \frac{\hat{p}^2}{2}+\frac{1}{2}\left(\epsilon^2+h\gamma^{x}_{1}\left(1+\frac{\gamma^{y}}{h}\right)\cos\Omega t\right)\hat{q}^2+\frac{N h}{2}
      ,
\end{equation}
%%%
where
%%%
\begin{equation}
      \label{EnergyScale}
            \epsilon = -h\sqrt{\left(1+\frac{\gamma^{y}}{h}\right)\left(1+\frac{\gamma^{x}_{0}}{h}\right)}
\end{equation}
%%%
is the characteristic energy scale of the system in the absence of driving $(\gamma^{x}_{1}=0)$.
The undriven system exhibits a critical behavior which is related to softening of the collective excitation spectrum, i.e., when the system is close to the critical point ($\gamma^{y}\rightarrow -h$),
the system exhibits a gapless excitation above the ground state 
($\epsilon\rightarrow 0$). Therefore, in the region $\left|\gamma^{x}_{0}\right|<-h<\gamma^{y}$, the symmetric phase became unstable. Interestingly, in the driven system the situation can be highly nontrivial 
as a consequence of mechanisms such as parametric resonance and parametric stabilization, which are characteristic of the parametrically-driven harmonic oscillator \cite{Arnold}. As a consequence, one can tune up conveniently the parameters close to resonance in order to manipulate the stability of the system, i.e., to produce a change of phase  for parameters which are far away from the critical point of the undriven system.
%%%%%%%%%%%%
\subsection{Resonance conditions}
%%%%%%%%%%%%
In the thermodynamic limit, the undriven LMG model is characterized by a single collective excitation with energy $\epsilon$. For a parametric oscillator with fundamental frequency $\epsilon$, $m$-photon transitions occur when the condition
%%%
\begin{equation}
      \label{ResonanceCond}
            2\epsilon=m\Omega     
      ,
\end{equation}
%%%
with integer $m$ is satisfied \cite{PeanoDykman,Shirley,dittrich,Grifoni, Nori}.

In this paper we focus on the parameter regime $\delta^{(m)},\gamma_{0}^{x},\gamma^{y} \ll \Omega$, where $\delta^{(m)}=-h-\frac{m\Omega}{2}$ is the detuning from resonance, and Eq. \eqref{ResonanceCond} reads
%%%
\begin{equation}
      \label{ResonanceCondWeak}
            -h\approx \frac{m\Omega}{2}    
      .
\end{equation}
%%%
%%%%
%%%%
\begin{figure}
\centering
\includegraphics[width=1.02\linewidth]{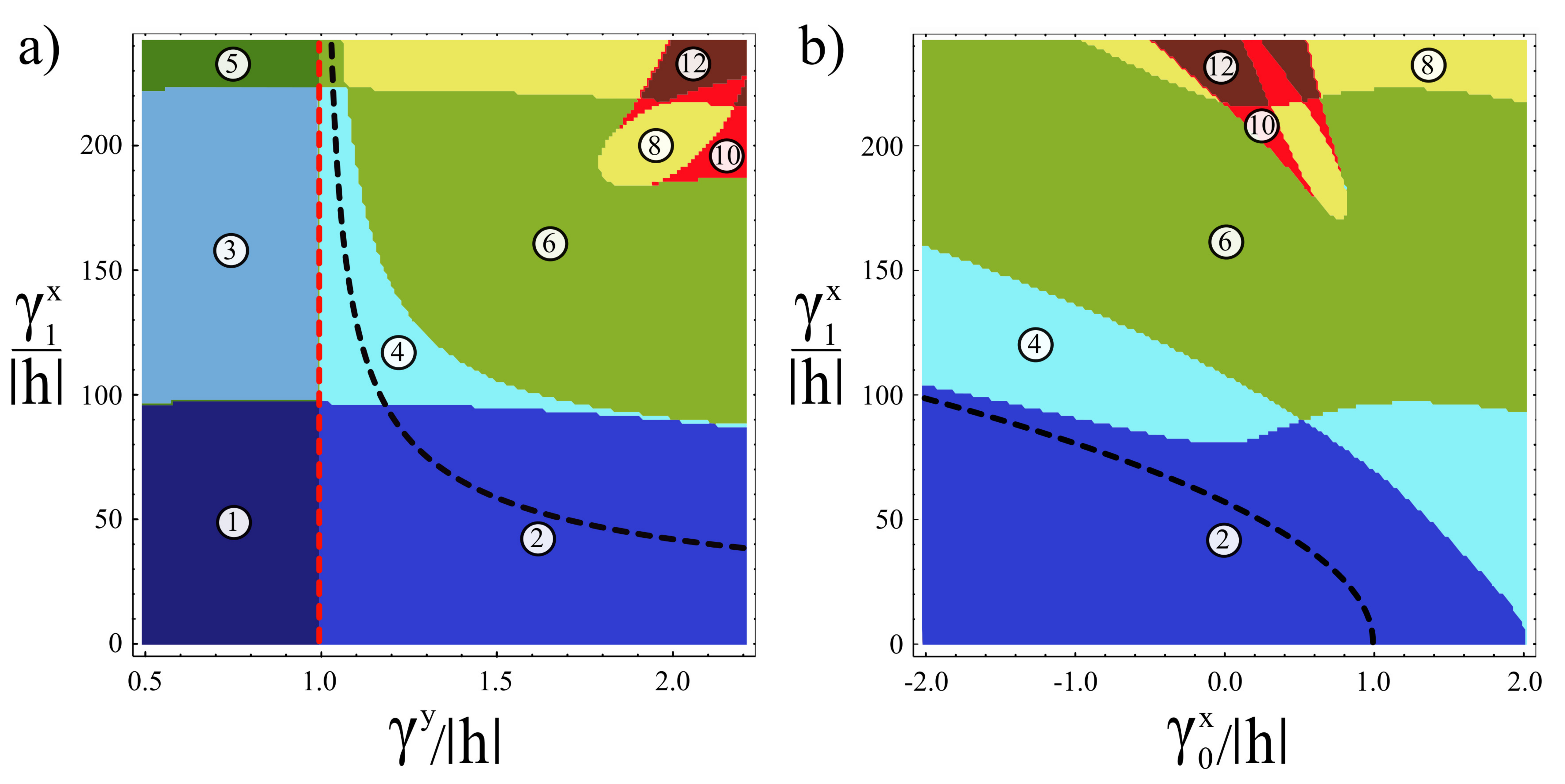}
\caption{(Color online)
  \label{Fig1}
  Phase diagram of the nonequilibrium QPT in the Lipkin-Meshkov-Glick model. The number of local minima of the quasienergy surface $E^{(0)}_{G}(Q,P)$ in the colored zones are indicated by the labels.
 (a) Depicts the phase diagram as a function of $\gamma^y$ and the driving amplitude $\gamma^{x}_1$ for 
 $\gamma^{x}_{0}/|h|=0.5$, and (b) as a function of $\gamma^{x}_0$ and $\gamma^{x}_1$ for $\gamma^{y}/|h|=2$. The dashed red (light gray) line in (a) resembles the second-order QPT in the undriven LMG model that occurs at $\gamma^{y}/|h|=1$, in the driven case, however, this line separates the regions with even and odd number of minima. The dashed black lines in (a) and (b) depict the transition of the symmetric phase $(Q,P)=(0,0)$ from a saddle point to a local maximum, and correspond to the contour $\lambda_2=0$.
 We consider the parameters $\Omega/|h|=40$ and $h/|h|=-1$.
}
\end{figure}
%%%%
%%%%
%%%%%%%%%%%%
\section{The rotating wave approximation and the effective Hamiltonian approach \label{SectionII}}
%%%%%%%%%%%% 
In this section we perform a study of the system based on the rotating wave approximation (RWA) \cite{PeanoDykman,Nori}.
In the limit $\delta^{(m)},\gamma_{0}^{x},\gamma^{y} \ll \Omega$, one can neglect the fast oscillations in the rotating frame, and obtain an effective time-independent Hamiltonian \cite{Bastidas3, Bastidas4, PeanoDykman, Nori}. Let us perform an unitary transformation of Hamiltonian Eq. \eqref{LMGDriven} into a convenient rotating frame via the unitary operator
%%%
\begin{equation}
      \label{UnitaryLMGRot}
            \hat{U}_{m}(t) =
            \exp \left(-i\Theta(t)J^{2}_{x}\right)\exp \left(- i \theta_m(t)J_z\right) 
     ,
\end{equation}
%%%
where $\Theta(t)=\frac{\gamma^{x}_{1}\sin\Omega t}{N\Omega}$ and $\theta_m(t) = \frac{m\Omega t}{2}$. In the rotating frame the dynamics is governed by the Hamiltonian
$\hat{H}_{m}(t)=\hat{U}^\dagger_{m}(t) \hat{\mathcal{H}} \hat{U}_{m}(t)$, where $\hat{\mathcal{H}}=\hat{H}(t)-i \frac{\partial}{\partial t}$ is the Floquet Hamiltonian \cite{Shirley}. The explicit form of this operator is given by (see Appendix \ref{AppendixA})
%%%
\begin{eqnarray}
      \label{HamiltonianRotFrameSI}
	    \hat{H}_{m}(t)&=& -\frac{h}{2}[(J_z+i\hat{\Lambda}^{m}_{1}(t))\hat{\mathcal{O}}^{m}_{1}(t)+h.c] -\frac{m\Omega}{2}J_z
	    \nonumber \\&&
	    +\frac{\gamma^y}{4N}[(J_z+i\hat{\Lambda}^{m}_{1}(t))^2\hat{\mathcal{O}}^{m}_{2}(t)+h.c]
	    \nonumber \\&&
            -\frac{\gamma^y}{2N}[J^{2}_{z}+(\hat{\Lambda}^{m}_{1}(t))^2]-\frac{\gamma^{x}_{0}}{N}(\hat{\Lambda}^{m}_{2}(t))^2
     .
\end{eqnarray}
%%%
We consider here the notation
%%%
\begin{align} 
	    \hat{\mathcal{O}}^{m}_{1}(t)&=\sum_{l=-\infty}^{\infty}\mathcal{J}_{l}\left[\frac{\gamma^{x}_{1}}{N\Omega}(2\hat{\Lambda}^{m}_{2}(t)+1)\right]e^{il\Omega t} , \label{FirstOperator}\\
	     \hat{\mathcal{O}}^{m}_{2}(t)&=\sum_{l=-\infty}^{\infty}\mathcal{J}_{l}\left[\frac{4\gamma^{x}_{1}}{N\Omega}(\hat{\Lambda}^{m}_{2}(t)+1)\right]e^{il\Omega t} 
       , \label{SecondOperator}
\end{align}
%%%
where $\mathcal{J}_{l}(z)$ is the  $l$th-order Bessel function \cite{Abramowitz},
and 
%%%
%%%
\begin{align}
	    \hat{\Lambda}^{m}_{1}(t)&= -J_y\cos\theta_m(t)-J_x\sin\theta_m(t) ,\label{RotJY}\\ 
	    \hat{\Lambda}^{m}_{2}(t)&= J_x\cos\theta_m(t)-J_y\sin\theta_m(t)
       . \label{RotJX}
\end{align}
%%%
The Hamiltonian Eq. \eqref{HamiltonianRotFrameSI} can be written in the form
%%%
\begin{equation}
      \label{LMGHamiltonianFourier}
             \hat{H}_{m}(t)=\sum_{n=-\infty}^{\infty} \hat{h}^{(m)}_{n}\exp{(i n \Omega t)}     
      .
\end{equation}
%%%
%%%%
%%%%
%%%%
\begin{figure}
    \includegraphics[width=0.85\linewidth]{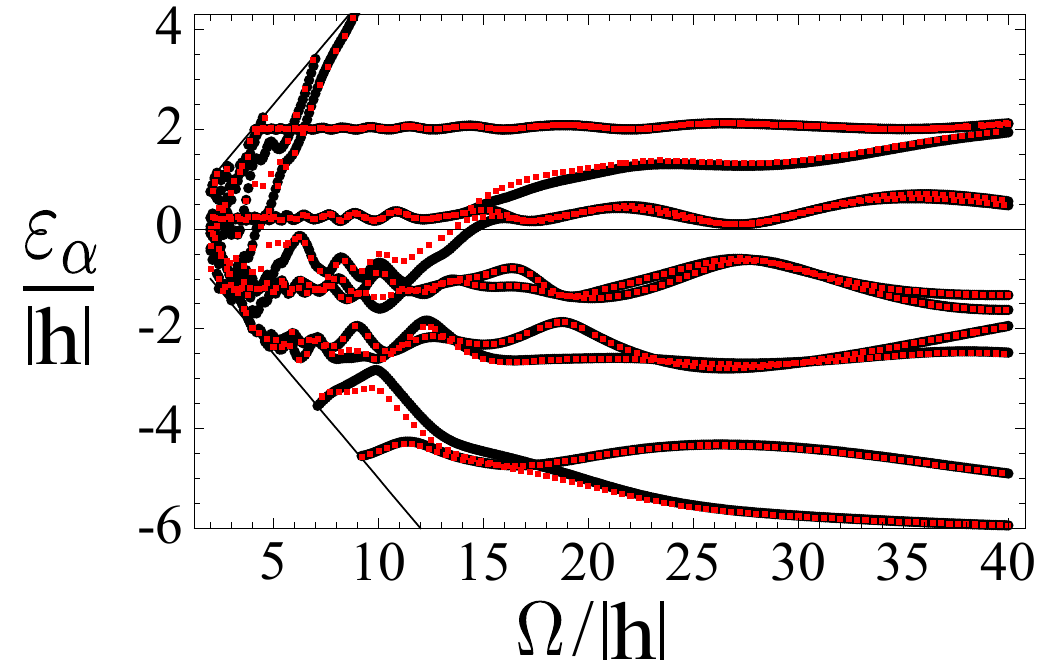}
    \caption{Comparison between the quasienergies obtained numerically (black curves) and the eigenvalues of the effective Hamiltonian $\hat{h}^{(0)}_{0}$ (red curves) as a function of the driving frequency $\Omega$. We consider a finite size system of $N=10$ particles and the parameters  $\frac{1}{|h|}(h,\gamma^x_0,\gamma^x_1,\gamma^y)=(-1,-1,210,2)$. The straight lines correspond to the boundaries $\varepsilon_{\alpha}=\pm\Omega/2$
    of the first Brillouin zone.}
    \label{Fig2M}
\end{figure}
%%%%
%%%%
Now we perform the RWA by neglecting all the time dependent terms of Eq. \eqref{LMGHamiltonianFourier}.
Under this approximation, the effective Hamiltonian $\hat{h}^{(m)}_{0}=\int^{T}_{0}\frac{dt}{T}\hat{H}_{m}(t)$ governs the dynamics in the rotating frame. Once one solves the eigenvalue problem
$\hat{h}^{(m)}_{0}|\widetilde{\Phi}_\alpha\rangle=\varepsilon_{\alpha}|\widetilde{\Phi}_\alpha\rangle$, the Floquet modes satisfying $\hat{\mathcal{H}}|\Phi_\alpha,t\rangle=\varepsilon_{\alpha}|\Phi_\alpha,t\rangle$ are given by 
$|\Phi_\alpha,t\rangle=\hat{U}_{m}(t)|\widetilde{\Phi}_\alpha\rangle$. The black curves in Fig. \ref{Fig2M} depicts the $N+1$ quasienergies $\varepsilon_{\alpha}$ obtained numerically for a system of $N=10$ particles. The red curves correspond to the eigenvalues of the effective Hamiltonian $\hat{h}^{(m)}_{0}$ for $m=0$. In  Fig. \ref{Fig2M} one can observe that RWA is a good approximation in the limit $\delta^{(0)},\gamma_{0}^{x},\gamma^{y} \ll \Omega$, where $\delta^{(0)}=-h$, as we have mentioned at the beginning of the section.

We next perform a bosonization procedure of $\hat{h}^{(m)}_{0}$ via the Holstein-Primakoff representation Eqs. \eqref{HPZ}, \eqref{HPMas}, and \eqref{HPMas}. In the bosonized version, the effective Hamiltonian is written in terms of the bosonic mode $b$. To investigate the criticality in the system we introduce a complex macroscopic displacement of order $\sqrt{N}$ for the bosonic operator as follows
%%%
\begin{equation}
      \label{MacroscopicDisplacement}
             b=c+\alpha\sqrt{N}     
      ,
\end{equation}
%%%
where $\alpha=(Q+iP)$ (Q and P are dimensionless parameters) and $c$ is a bosonic operator describing quantum fluctuations in the system. In the thermodynamic limit $N\rightarrow\infty$, we perform a series expansion of the effective Hamiltonian in powers of $\sqrt{N}$
%%%
\begin{equation}
      \label{SeriesExpEffHam}
            \hat{h}^{(m)}_{0}=\hat{h}^{(m)}_Q(c,c^{\dagger})+\sqrt{N}\ \hat{h}^{(m)}_{L}(c,c^{\dagger})+N E^{(m)}_{G}(Q,P)    
      ,
\end{equation}
%%%
where $\hat{h}^{(m)}_Q(c,c^{\dagger})$ is a quadratic bosonic Hamiltonian, $\hat{h}^{(m)}_{L}(c,c^{\dagger})$ contains linear
bosonic terms, and $E^{(m)}_{G}(Q,P)$ is the quasienergy landscape (QEL). The linear term $\hat{h}^{(m)}_{L}(c,c^{\dagger})$ of the Hamiltonian expansion Eq. \eqref{SeriesExpEffHam} vanishes for macroscopic displacements located at a critical point of the QEL, as the coefficients of the expansion are proportional to the first derivatives of $E^{(m)}_G(Q,P)$ with respect to $Q$ and $P$. Correspondingly, the quadratic term $\hat{h}^{(m)}_{Q}$ contains the geometric information of the principal curvatures in the neighborhood of a local minimum as it contains second derivatives of the QEL with respect to $Q$ and $P$. Interestingly, such principal curvatures are nothing but the energies of the collective excitations characterizing the QPT \cite{Hayn}. Therefore, when a stable configuration of the QEL landscape exhibits a transition into a saddle point or a local maximum, there is a softening of a collective excitation, that is related---from the geometrical point of view, to an infinite curvature radius in the 
neighborhood of the critical point. Beyond the mean field description, the quantum corrections can be investigated by means of the continuous unitary transformations method \cite{VidalFSExp}.
%%%%%%%%%%%%
\subsection{The effective Hamiltonian for the $m=0$ case}
%%%%%%%%%%%%
Instead of performing an abstract general theory for the effective Hamiltonian related to general $m$-photon resonances \cite{PeanoThorwart, LeytonThorwart}, we focus here on an illustrative particular case that contains the more relevant information, i.e., we consider the case $m=0$.
In this case, the effective Hamiltonian reads
%%%
\begin{widetext}
      \begin{eqnarray}
            \label{EffectiveHam0}
                  \hat{h}^{(0)}_{0} =\left[-\frac{h}{2}(J_z - i J_y) \mathcal{J}_0\left[\frac{\gamma^{x}_{1}}{N\Omega}(2J_x+1)\right]
                   +\frac{\gamma^{y}}{4N}(J_z-iJ_y)^2 \mathcal{J}_0\left[\frac{4\gamma^{x}_{1}}{N\Omega}(J_x+1)\right]
                           +  h.c \right]-\frac{\gamma^{y}}{2N}(J_z^2+J_y^2) -\frac{\gamma^{x}_{0}}{N} J_x^2 \
            .               
      \end{eqnarray}
\end{widetext}
%%%
In the thermodynamic limit $N\rightarrow \infty$, we expand the Holstein-Primakoff representation Eqs. \eqref{HPZ}, \eqref{HPMas}, and \eqref{HPMas} with respect to the complex macroscopic displacement Eq. (\ref{MacroscopicDisplacement}). We next consider the scaled angular momentum operators
%%%
\begin{align}
      \hat{X}_{1} &=\frac{J_{x}}{N}= Q \sqrt{1-\left|\alpha\right|^2}         ,  \label{eq:holprimx} \\
      \hat{X}_{2} &=\frac{J_{y}}{N}= P \sqrt{1-\left|\alpha\right|^2}       ,   \label{eq:holprimy}   \\
      \hat{X}_{3} &=\frac{J_{z}}{N}=\left( \left|\alpha\right|^2-\frac{1}{2}\right) ,   \label{eq:holprimz}
\end{align}
%%%
that satisfy $[\hat{X}_{i},\hat{X}_{j}]=0$ in the thermodynamic limit. Therefore, in this limit, the scaled angular momentum operators become $c$-number variables.
Eqs. \eqref{eq:holprimx}, \eqref{eq:holprimy}, and \eqref{eq:holprimz} describe a mapping from the coordinates $(Q,P)$ onto the Bloch sphere, because the norm of the vector $\textbf{R}=\left(X_1,X_2,X_3\right)$ has constant length $\rVert\textbf{R}\lVert=1/2$. From Eqs. \eqref{eq:holprimx} and \eqref{eq:holprimy} it follows that $\left|\alpha\right|^2 \le 1$. 
Furthermore, one can see from these relations that all points $(Q,P)$ with $\left|\alpha\right|^2= 1$ correspond to  $X_3=1/2$. Therefore, all points of the boundary $\left|\alpha\right|^2= 1$ are mapped into the north pole of the Bloch sphere. For the interior points $\left|\alpha\right|^2< 1$, the transformation  is bijective, e.g., the origin $(Q,P)=(0,0)$ is mapped into the south pole $X_3=-1/2$.
\\
By replacing Eqs. \eqref{eq:holprimx},\eqref{eq:holprimy}, and \eqref{eq:holprimz} into the effective Hamiltonian Eq. \eqref{EffectiveHam0} we obtain the QEL for the $m=0$ case
%%%
\begin{widetext}
      \begin{eqnarray}
            \label{eq:eg}
                  E_G^{(0)}(Q,P) &=& -h \left( \left|\alpha\right|^2-\frac{1}{2}\right) \mathcal{J}_0\left[\frac{2\gamma_1^x}{\Omega}                                                                  Q\sqrt{1-\left|\alpha\right|^2}\right] 
                 -\frac{\gamma^y}{2}\left[ \left( \left|\alpha\right|^2-\frac{1}{2}\right)^2+\left(1-\left|\alpha\right|^2\right) P^2\right]
                  \nonumber \\&&
                 +\frac{\gamma^y}{2}\left[\left( \left|\alpha\right|^2-\frac{1}{2}\right)^2-\left(1-\left|\alpha\right|^2\right)P^2\right]                                             \mathcal{J}_0\left[\frac{4\gamma_1^x}{\Omega }Q\sqrt{1-\left|\alpha\right|^2}\right]
                 -\gamma_0^x \left(1-\left|\alpha\right|^2\right)Q^2 
           .
      \end{eqnarray}
\end{widetext}
%%%
In the undriven case, a classification of the quantum phases of the LMG Hamiltonian is usually performed by studying
the global minima of the energy landscape \cite{RibeiroESQPT}. However, in this work we interpret the local minima of the QEL
as metastable phases of the driven system. These metastable
states are related to the phenomenon of parametric stabilization
\cite{Arnold}. Since these are separated from the global
minima by macroscopic displacements, we expect transitions
to be suppressed, such that
the corresponding values of the order parameters are observable.
%%%%
%%%%
\begin{figure}
\vspace{0.4cm}
  \begin{minipage}[b]{0.81\linewidth}
    \begin{overpic}[width=\linewidth]{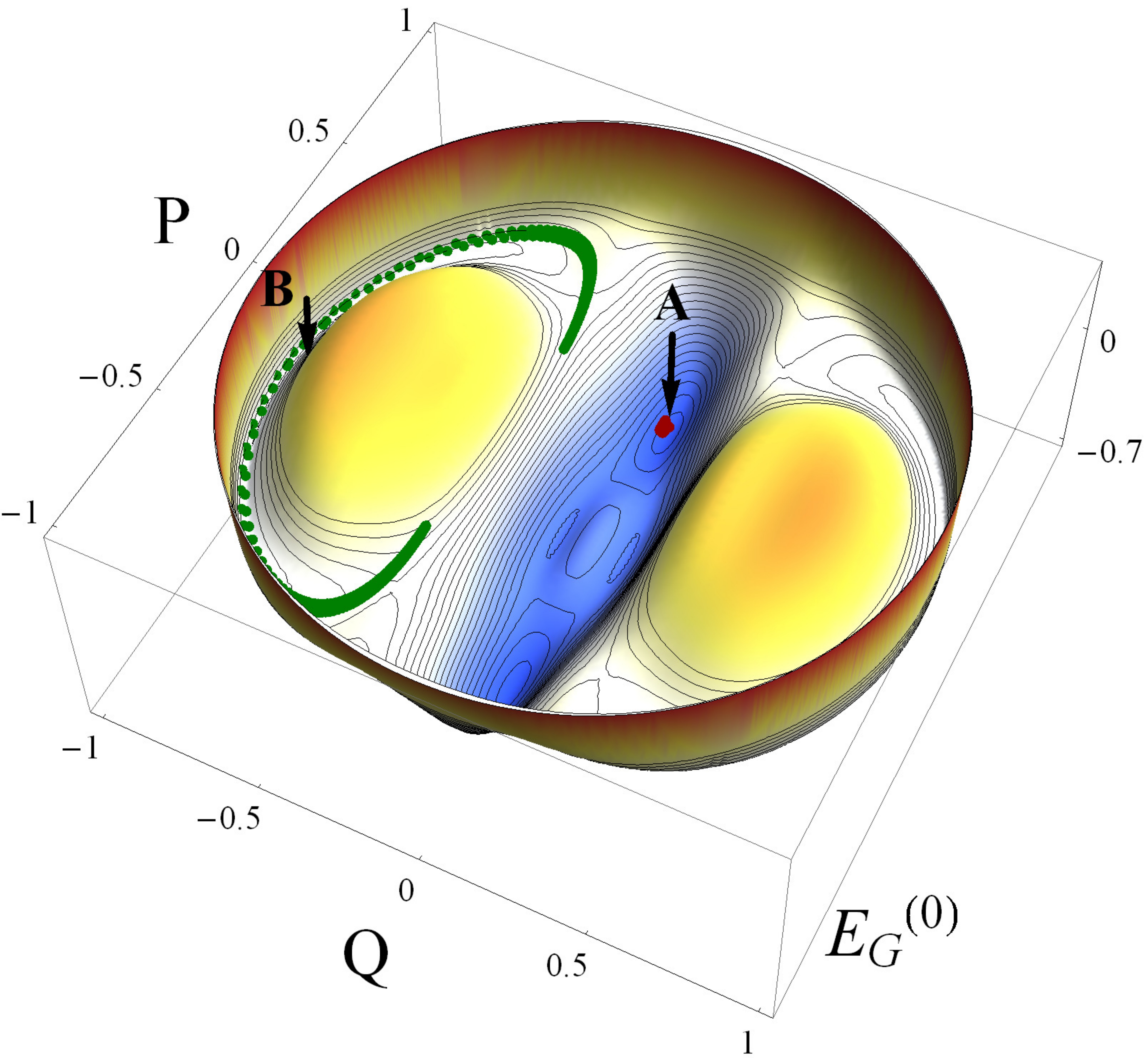}
    \put(-2,90){ \textbf a)}
    \end{overpic}
  \end{minipage}
  \begin{minipage}[b]{0.81\linewidth}
    \begin{overpic}[width=\linewidth]{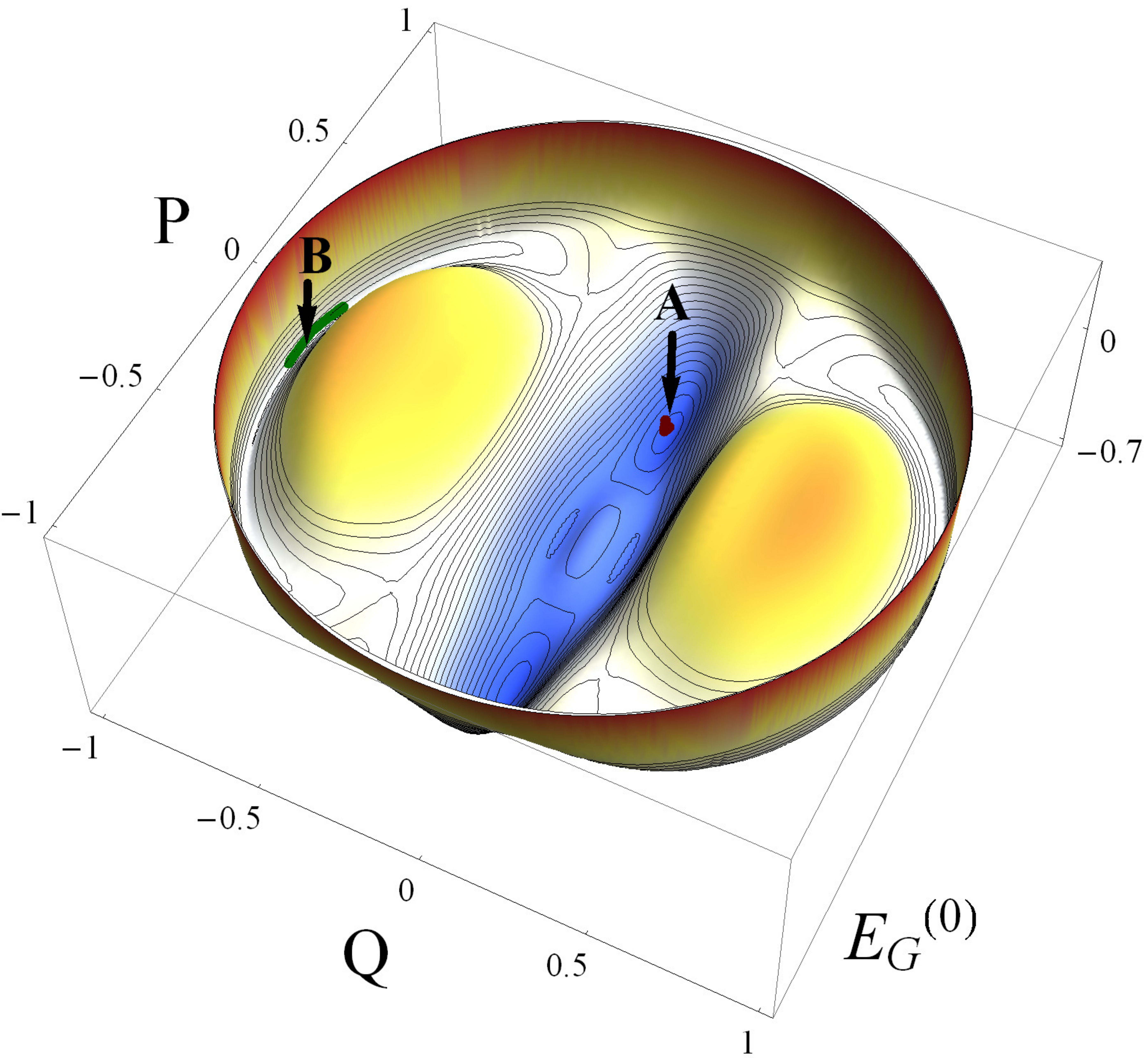}
    \put(-2,90){ \textbf b)}
    \end{overpic}
  \end{minipage}
    \caption{Quasienergy landscape $ E_G^{(0)}(Q,P)$ for the parameters $\frac{1}{|h|}(h,\gamma^x_0,\gamma^x_1,\gamma^y)=(-1,-1,210,2)$. In the undriven system, these parameters correspond to the symmetry-broken phase. The quantum evolution for $N=100$ particles within one period is calculated when the system is initialized in a spin coherent state. 
    (a) Depicts the quantum evolution of the operators in the laboratory frame, and (b) the effective evolution in the rotating frame.
    The green (light gray) line on the surface depicts the evolution of an initial wave packet centered at minimum $B$ and the red line shows the corresponding evolution for a wave packet initially centered at $A$. 
    We consider the parameters $m=0$ and $\Omega/|h|=40$.}
    \label{Fig2}
\end{figure}
%%%%
%%%%
%%%%
%%%%
\begin{figure}
\vspace{0.4cm}
  \begin{minipage}[b]{0.49\linewidth}
    \begin{overpic}[width=\linewidth]{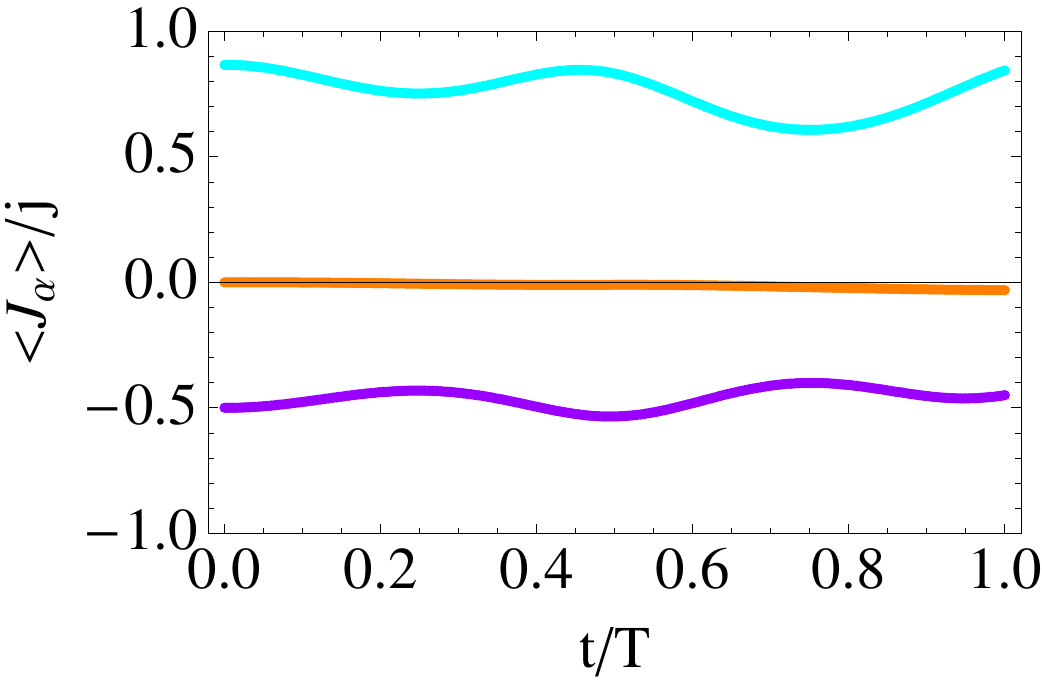}
    \put(-2,60){ \textbf a)}
    \end{overpic}
  \end{minipage}
  \begin{minipage}[b]{0.49\linewidth}
    \begin{overpic}[width=\linewidth]{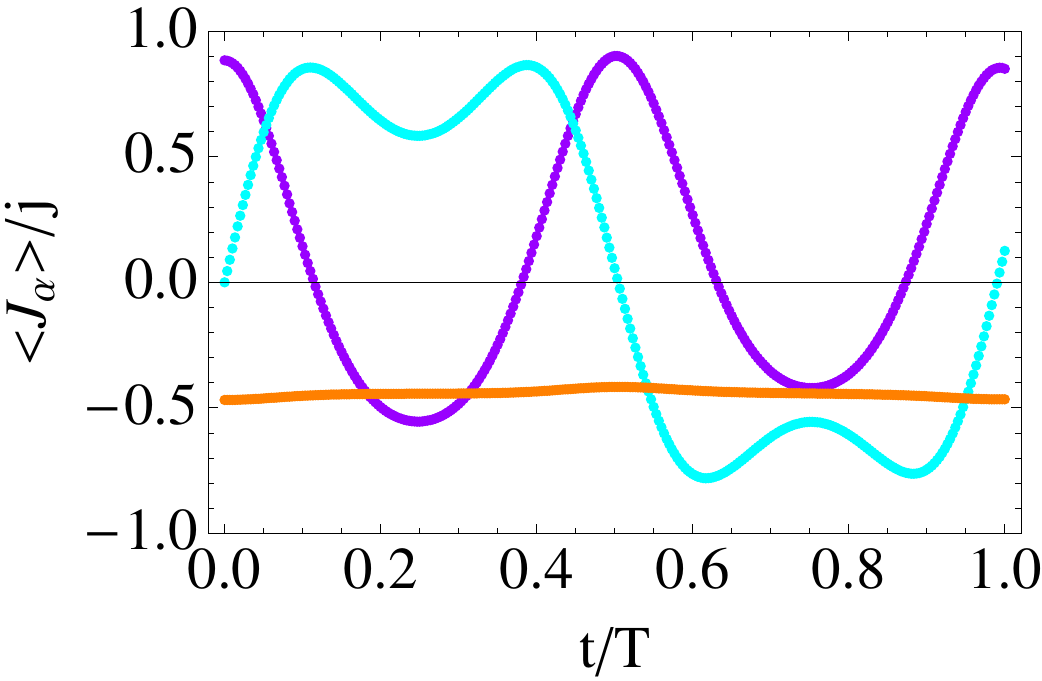}
    \put(-2,60){ \textbf d)}
    \end{overpic}
  \end{minipage}
  \begin{minipage}[b]{0.49\linewidth}
    \begin{overpic}[width=\linewidth]{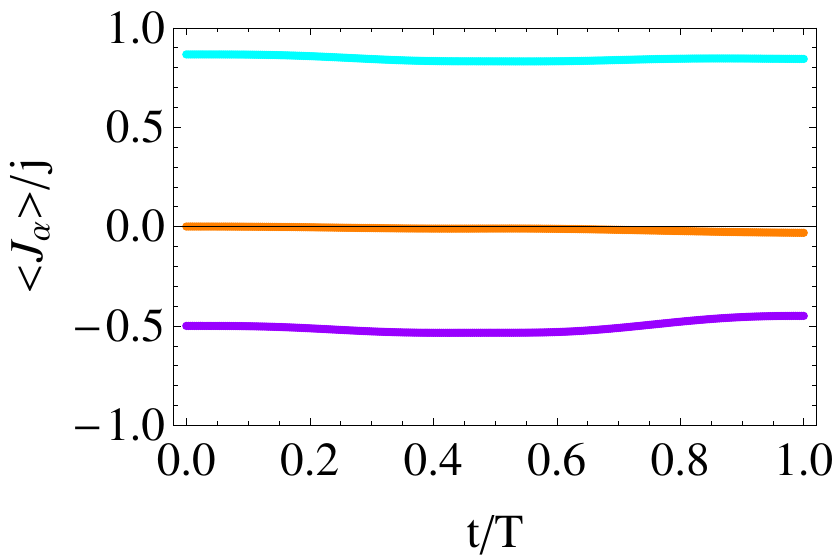}
    \put(-2,60){ \textbf b)}
    \end{overpic}
  \end{minipage}
  \begin{minipage}[b]{0.49\linewidth}
    \begin{overpic}[width=\linewidth]{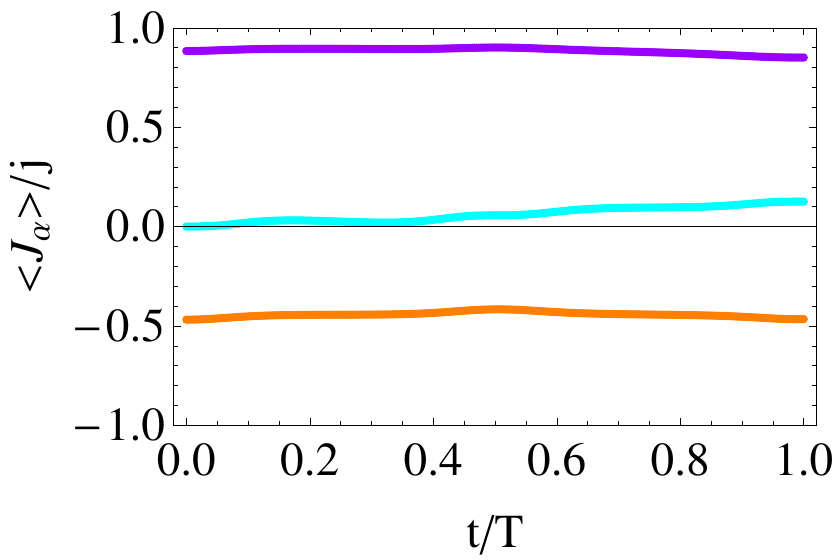}
    \put(-2,60){ \textbf e)}
    \end{overpic}
  \end{minipage}
    \begin{minipage}[b]{0.49\linewidth}
    \begin{overpic}[width=\linewidth]{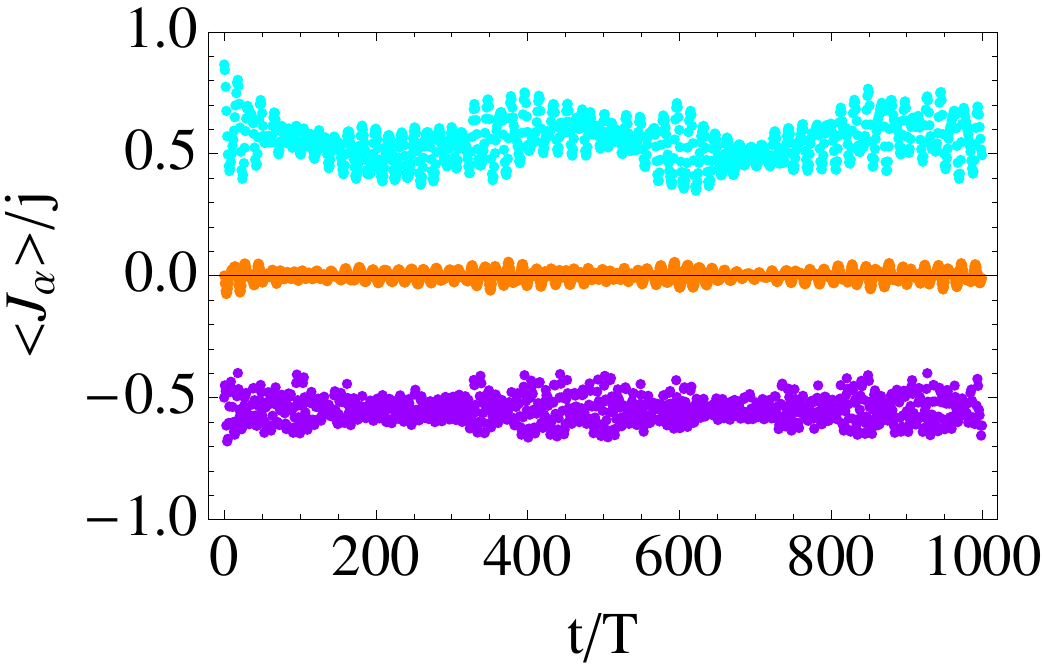}
    \put(-2,60){ \textbf c)}
    \end{overpic}
  \end{minipage}
  \begin{minipage}[b]{0.49\linewidth}
    \begin{overpic}[width=\linewidth]{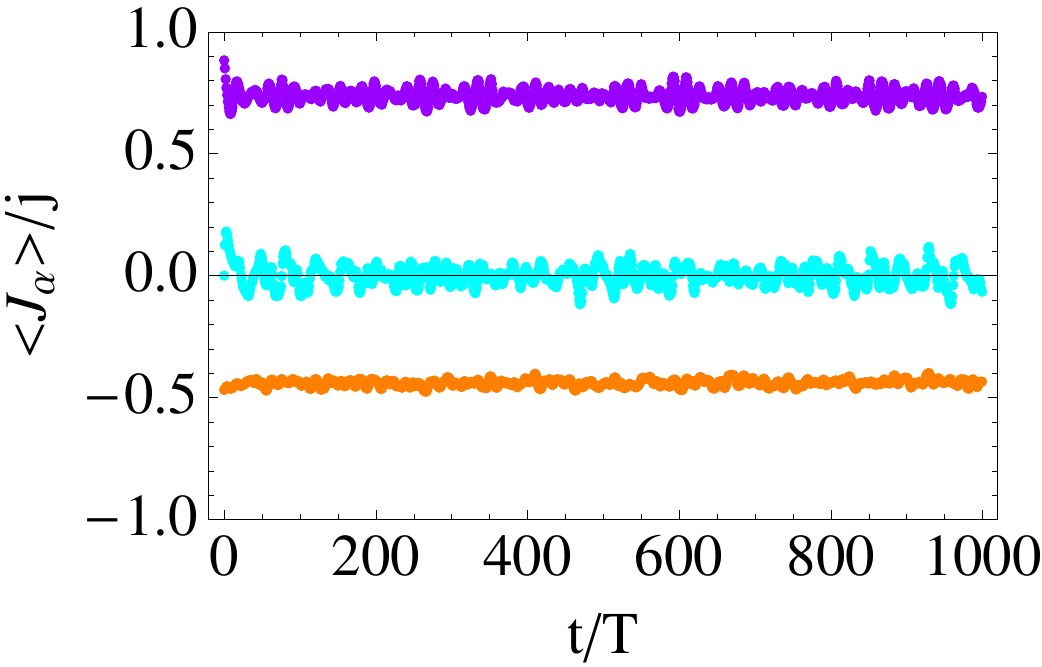}
    \put(-2,60){ \textbf f)}
    \end{overpic}
  \end{minipage}
  \caption{Quantum evolution of the observables for finite size $N=100$ and parameters $\frac{1}{|h|}(h,\gamma^x_0,\gamma^x_1,\gamma^y)=(-1,-1,210,2)$. The expectation values $\langle J_x\rangle/j,\langle J_y\rangle/j$ and $\langle J_z\rangle/j$ are depicted by the orange, cyan and magenta curves respectively.
  For an initial wave packet centered at the minimum $A$ in Fig. \ref{Fig2}: 
  (a) depicts the quantum evolution within one period, (b) the effective evolution in the rotating frame, and (c) depicts the stroboscopic dynamics. 
  Correspondingly, when the system is initialized in a wave packet centered at the minimum $B$ in Fig. \ref{Fig2}:
  (d) depicts the quantum evolution within one period, (e) the effective evolution, and (e) the stroboscopic evolution. We consider the parameters $m=0$ and $\Omega/|h|=40$.}
  \label{Fig3}
\end{figure}
%%%%
%%%%
This possibility is reinforced when one recalls
that $ E_G^{(0)}(Q,P)$  does not have the same thermodynamic significance
as the lowest actual energy \cite{Bastidas3}.
Therefore, in this work we consider the number of minima of the QEL as a criterion to establish a new phase diagram. 

One can consider an alternative description in terms of the solution of the equations of motion for the expectation values of the angular momentum operators. In the thermodynamic limit, it is possible to factorize
the averages of operator products \cite{Parkins1}. Therefore, one obtains a closed set of semiclassical equations of motion. In this context, the critical points of the QEL in Eq. \eqref{eq:eg} would correspond to the fixed points of the Poincar\'{e} map \cite{Arnold}. Correspondingly, the minina of the QEL are the stable periodic trajectories of the Hamiltonian flow \cite{Arnold}.

Fig.\ref{Fig1} (a) depicts the phase diagram as a function of $\gamma^{y}$ and $\gamma^{x}_{1}$ , for fixed $\gamma^{x}_{0}/|h|=0.5$, and Fig.\ref{Fig1} (b) as a function of $\gamma^{x}_{0}$ and $\gamma^{x}_{1}$, for fixed $\gamma^{y}/|h|=2$. In these  phase diagrams, we see many regions corresponding to different number of minima of the QEL. Additionally, the phase diagrams show the appearance of many novel metastable phases, which are separated from each other by boundary lines, whose crossings correspond to nonequilibrium multicritical points. By crossing the line $\gamma^y=-h$ in Fig.\ref{Fig1}, the single global minimum at $(Q,P)=(0,0)$ splits into two macroscopically separated global minima thus resembling the second order QPT known from the time-independent model. Interestingly, regions with even and odd number of minima are characterized by the existence of two and one global minima respectively. We can study the stability of the global minimum at $(Q,P)=(0,0)$ analytically by computing 
the Jacobian-matrix at the origin of the QEL and its eigenvalues
%%%	
\begin{align}
      \label{eq:eigenwerte}
	    \lambda_1&=-2\left(h+\gamma^y \right), \\ 
            \lambda_2&=-2h-2\gamma^x_0-\left(h+\gamma^y \right) \left( \frac{\gamma^x_1}{\Omega} \right)^2 
      .
\end{align} 
%%%
Both eigenvalues are positive in the region $\gamma_y< -h$ (the contours $\lambda_1=0$ and $\lambda_2=0$ are depicted as red (light gray) and black dashed lines respectively in Fig.\ref{Fig1}).
In the region $\gamma_y>-h$ and $\gamma^{x}_{1} < \Omega \sqrt{\left(\frac{2h+2\gamma^{x}_{0}}{-h-\gamma^{y}}\right)} $ (region between the dashed curves in Fig.\ref{Fig1} (a)), $\lambda_1$ is negative and $\lambda_2$ is positive. Furthermore $\lambda_1$ and $\lambda_2$ are negative for $\gamma^{x}_{1} > \Omega \sqrt{\left(\frac{2h+2\gamma^{x}_{0}}{-h-\gamma^{y}}\right)}$, and therefore, by crossing the curve $ \lambda_2=0$, the saddle point at the origin becomes a local maximum. Consequently, in the region $\gamma_y>-h$ the single central minimum splits up in two global minima. The phase diagram depicted in Fig.\ref{Fig1} (b) is characterized by $\lambda_1<0$, and therefore, the dashed line corresponds to the boundary between the regions below and above the level curve $\lambda_2=0$, where the origin is a saddle point and a maximum respectively. 

An example for the QEL is given in Fig. \ref{Fig2}, where the parameters are so chosen, that the undriven system ($\gamma^{y}_{1}=0$) is in the symmetry-broken phase. As the QEL corresponds to the  energy landscape in the undriven case, it exhibits two global minima corresponding to macroscopically separated states degenerate in energy.  In the driven system, apart from the  two global minima characteristic of the undriven system (in Fig. \ref{Fig2} denoted with A), new characteristics of the QEL appear, e.g.  two local minima (denoted with B). These new local minima can be interpreted to be novel metastable states, which are strongly related to the quantum evolution of the system as we describe in the next section.

%%%%%%%%%%%%
\subsection{Quantum evolution}
%%%%%%%%%%%%
In this section we investigate the quantum evolution when the system is initially prepared in a spin coherent state \cite{Arecchi} centered at a local minimum of the QEL. Spin coherent states have minimal uncertainty and are the closest ones to a classical angular momentum state. To describe geometrically the quantum evolution, we parametrize the Bloch sphere using spherical coordinates and express the angular momentum components in terms of the azimuthal $(\phi)$ and  polar $(\theta)$ angles as follows:
%%%
\begin{align}
      \label{eq:} 
	    X_1 &= \frac{1}{2} \sin\theta \cos\phi    , \\
	    X_2 &= \frac{1}{2} \sin \theta  \sin \phi                   ,             \\
	    X_3 &= - \frac{1}{2} \cos \theta          
     .  
\end{align}
%%%
By replacing this set of equations into Eqs. \eqref{eq:holprimx}-\eqref{eq:holprimz} we find a relation between the $(\theta,\phi)$ and $(Q,P)$ coordinate systems
%%%
\begin{align}
            \theta &= \pi-\arccos\left[2\left(Q^2+P^2\right)-1 \right] , \label{eq:transformation1}\\
            \phi   &= \arccos \left[\frac{2Q}{\sin\left(\theta\right)}\sqrt{1-Q^2-P^2}\right] \label{eq:transformation2}
      .
\end{align}
%%%
With the angular coordinates $\phi$ and $\theta$, we can represent the spin coherent state $\left|\phi,\theta \right> $  by using 
%%%
\begin{equation}
      \label{SpinCohState}
            \left| \phi,\theta \right> =(1+\left|\tau\right|^2)^{-j} \exp[\tau J_{+}] \left|j,-j \right> 
      ,
\end{equation}
%%%
with $\tau=e^{-i\phi}\tan \frac{\theta}{2}$. This procedure  describes a mapping from a point of the QEL onto the set of spin coherent states.

To study the quantum evolution we consider a  system consisting of $N=100$ particles. For a finite size $N$, the numerical problem consists in the solution of $N+1$ coupled ordinary differential equations. After the numerical integration of the Schr\"odinger equation, we construct the evolution operator $\hat{U}(t,0)$, which allows us to calculate the state of the system at any time $t>0$: $\left|\Psi, t\right> = \hat{U}(t,0) \left|\phi,\theta \right>$, providing that we prepare initially the system in a spin coherent state $\left|\Psi,0 \right> =\left|\phi,\theta \right>$. We now proceed to calculate the normalized expectation values of the spin components in the state $\left|\Psi, t\right>$
%%%
\begin{equation}
      \label{ExpectValuesLab}
            \langle J_{\alpha}\rangle = \langle \Psi, t|J_{\alpha}|\Psi, t\rangle
      ,
\end{equation}
%%%
with $\alpha \in \{x,y,z\}$.

Fig. \ref{Fig3} (a) depicts the continuous time evolution within one period of the driving $T=2\pi/\Omega$ of an initial wave packet centered at the minimum $A$ in Fig. \ref{Fig2}. Similarly, Fig. \ref{Fig3} (d) shows  the inter-period  dynamics  when the initial wave packet is centered at the minimum $B$ in Fig. 
\ref{Fig2}.
To  obtain a better geometrical picture of the quantum evolution we represent the mean values of the angular momentum components by means of the coordinates $Q$ and $P$. Therefore, we project the expectation values onto the Bloch sphere by calculating the angles $\phi$ and $\theta$ and then solving Eqs. \eqref{eq:transformation1} and  \eqref{eq:transformation2} for  $Q$ and $P$.
The result is shown in Fig. \ref{Fig2} (a) for $N=100$ particles, where the time evolution $\langle J_{\alpha}\rangle$ for initial wave packets centered in minima $A$ and $B$ is depicted by the red and green (light gray) curves, respectively. The trajectory initialized in A is strongly trapped within the minimum, whereas the other trajectory  exhibits higher oscillations around the initial state. For finite size $N \gg 1$ the trajectories take place approximately over the surface of the QEL, i.e., the mean value of the spin evolves in average along points with equal  values of $E_G^{(0)}(Q,P)$. This behavior is connected to the fact that the eigenvalues of the effective Hamiltonian---from which the QEL is derived, correspond to the  quasienergies of the system and the average value of the quasienergy is conserved in a time-periodic system.

The state in the laboratory frame $|\Psi,t\rangle$ and the state in the rotating frame $|\Psi_{m} ,t\rangle$ --- in which the QEL is derived, are connected via
%%%
\begin{align}
      \label{stateRot}
            |\Psi,t\rangle &= \hat U _{m}(t) |\Psi_{m} ,t\rangle  \nonumber \\  
                           & \approx \hat U _{m}(t) \hat e^{-i \hat h_0^{(m)} t} |\Psi,0\rangle
     ,
\end{align}
%%% 
where $\hat{U}_{\text{rot}}(t,0) = \hat e^{-i \hat{h}_0^{(m)} t}$ denotes the propagator in the rotating frame and $|\Psi_{m} ,0\rangle=|\Psi,0\rangle$ as a consequence of Eq. \eqref{UnitaryLMGRot}. The propagators in the laboratory and rotating frame are related as $\hat{U} (t,0)\approx \hat{U} _m(t)  e^{-i \hat{h}_0^{(m)} t}$.

For a better understanding of the quasienergy landscape, let us consider an effective quantum evolution
in the rotating frame. We consider the expectation value
%%%
\begin{equation}
      \label{ExpectValuesRot}
            \langle J_{\alpha}\rangle^{(0)} = \langle \Psi_{m}, t|J_{\alpha}|\Psi_{m}, t\rangle
      ,
\end{equation}
%%%
with $\alpha \in \{x,y,z\}$, and $|\Psi_{m}, t\rangle=\hat{U}^{\dagger}_{m}(t)|\Psi,t\rangle$.

Fig. \ref{Fig3} (b) and (e) show the effective evolution $\langle J_{\alpha}\rangle^{(0)}/j$ within one period of an initial wave packet centered at the minima $A$  and $B$ in Fig. \ref{Fig2}, respectively. Similarly to the procedure used previously for the evolution in the laboratory frame, Fig. \ref{Fig2} (b) depict the effective evolution of observables in the QEL. In contrast to the evolution in the laboratory frame, for the same $N=100$ particles, the effective evolution exhibits higher localization around the initial condition. Eq. \eqref{ExpectValuesRot} can be interpreted alternatively as the expectation value of the rotating operator 
%%%
\begin{equation}
      \label{RotatingOp}
            J^{\text{rot}}_{\alpha}(t)=\hat{U}_m(t)J_{\alpha}\hat{U}^{\dagger}_m(t)
\end{equation}
%%%
in the state $|\Psi, t\rangle$.

Next we consider periodic snapshots of the evolution of observables, this generates a discrete evolution in time, commonly referred to as stroboscopic quantum  evolution.
There is a very interesting relation between the stroboscopic quantum evolution and the parity operator Eq. \eqref{Parity} that can be established using the relation
%%%
\begin{equation}
      \label{ParityCondition}
             \hat{U} _m(t_r)=\left[\hat{\Pi}\exp\left(i\frac{\pi N}{2}\right)\right]^{\frac{m r}{2}}               
      ,
\end{equation}
%%%
for $t_r= \frac {2 r \pi}\Omega$ with integer $r$.
As a consequence of this, for $m=0$ we have $ \hat{U} _m(t_r)=\hat{\mathbbm{1}}$, which implies that the states in the laboratory frame and in the rotating frame are identical for these times. Thus,  the stroboscopic time evolution is governed entirely by $\hat h_0^{(0)}$. The stroboscopic long-time evolution of the observables is displayed in Fig. \ref{Fig3}(c). Here,  the quantum evolution is strongly trapped in the neighborhood of the minimum $A$ in Fig. \ref{Fig2}. A similar situation occurs in Fig. \ref{Fig3}(f) for a wave packet initially centered at the minimum $B$ in Fig. \ref{Fig2}.  By calculating the quantum evolution  with different system sizes, one finds that the quantum fluctuations of the trajectories decrease with number of particles, and therefore, in the thermodynamic limit, the time evolution has to be constrained to the QEL. Because the time evolution is connected with the geometrical features of the QEL, it is justified to use it as a background to define the existence of new 
metastable phases.

%%%%
\section{Conclusions\label{SectionIII}}
%%%%
We have investigated the driving-induced QPT in the LMG model. We show that under the effect of an external driving the system exhibits a multistable character with no analogue in equilibrium systems. In particular, the novel quantum phases correspond to local minima of the QLE landscape. To understand the nature of the nonequilibrium metastable states, we study the quantum evolution for finite size $N$ when the system is initially prepared  in a coherent state centered at one particular local minimum of the QEL. We find that the system is dynamically trapped in the neighborhood of the initial state, and the quantum evolution is related to the local curvature of the chosen minima, i.e., for higher curvature the trapping effect is stronger. Our approach opens a new window in the understanding of driving-induced criticality, in particular, the effective 
Hamiltonian Eq. \eqref{EffectiveHam0} contains effective interactions  which are absent in equilibrium, in this sense, the effective Hamiltonian can be considered as a quantum simulator. 

Ref. \cite{Parkins2} reports on the relation between the uniaxial LMG model and the Dicke model in the dispersive regime. A possible experimental realization of our model can be based on a Bose-Einstein condensate coupled to an optical cavity. Such a setup allows for a simulation of superradiant QPT in the Dicke model \cite{Esslinger, Mottl}. In the experiment, the frequency of the cavity is the dominant energy scale. In other words, the experimental parameters allows for the investigation of the dispersive regime. In this regime, the Hamiltonian Eq. \eqref{LMGDriven} is realized by considering $-h$ 
as twice the recoil energy of the condensate, and $\gamma^{y}=0$. The time dependent inter-particle interaction $\gamma^{x}(t)$ could be controlled externally by varying the intensity of the pump laser as a function of time about a static value \cite{Bastidas3, Parkins2}.

Surprisingly, despite of the close relation between the LMG model and the Dicke model, in contrast to our description of the nonequilibrium QPT in the Dicke model \cite{Bastidas3}, the order of the phase transitions in the LMG model does not change under the effect of driving.
In the case of undriven QPTs, the critical behavior is related to nonanalyticities of the ground state. Under the effect of an external driving, however, the system will experience transitions to excited states, even if it is initially prepared in the ground state. In this sense, the global minima as well as the local minima of the QEL play the role of driving-induced states of matter. 

%%%%%
\begin{acknowledgments}
%%%%%
The authors gratefully acknowledge financial support from the DAAD and DFG Grants BR $1528/7-1$, $1528/8-1$, SFB $910$, GRK $1558$, and SCHA $1646/2-1$.
\end{acknowledgments}
%%%%%
%
\appendix
%
%%%%%%%%
\section{Derivation of the Hamiltonian in the rotating frame $\hat{H}_{m}(t)$ \label{AppendixA}}
%%%%%%%% 
Our aim in this appendix is to derive explicitly the Hamiltonian Eq. \eqref{HamiltonianRotFrameSI}.
Let us consider first a transformation into the interaction picture by means of the unitary operator
$\hat{U}_{I}(t)=\exp \left(-i\Theta(t)J^{2}_{x}\right)$, where $\Theta(t)=\frac{\gamma^{x}_{1}\sin\Omega t}{N\Omega}$. In the interaction picture, the state of the system is given by $\rvert \Psi_{I},t\rangle=\hat{U}^{\dagger}_{I}(t)\rvert \Psi,t\rangle$, where $\rvert \Psi,t\rangle$ is the state in the Schr\"odinger picture. Correspondingly, the Hamiltonian in the interaction picture reads
%%%
\begin{align}
      \label{HamiltonianInteraction}
            \hat{H}_{I}(t)&=\hat{U}^{\dagger}_{I}(t)\left(\hat{H}(t)-i \frac{\partial}{\partial t}\right)\hat{U}_{I}(t) \nonumber
            \\
            &= -\frac{h}{2}[(J_z-iJ_y)\hat{\mathcal{O}}_{1}(t)+h.c] 
	    \nonumber \\&
	    +\frac{\gamma^y}{4N}[(J_z-iJ_y)^2\hat{\mathcal{O}}_{2}(t)+h.c]
	    \nonumber \\&
            -\frac{\gamma^y}{2N}[J^{2}_{z}+J_{y}^{2}]-\frac{\gamma^{x}_{0}}{N}J_{x}^{2} \
      ,
\end{align}
%%%
where 
%%%
\begin{align}
            \hat{\mathcal{O}}_{1}(t)&=\exp \left[i\Theta(t)(2J_{x}+1)\right]
           \nonumber \\ 
            &= \sum_{l=-\infty}^{\infty}\mathcal{J}_{l}\left[\frac{\gamma^{x}_{1}}{N\Omega}(2J_{x}+1)\right]e^{il\Omega t} ,
              \label{Exponential1} \\
            \hat{\mathcal{O}}_{2}(t)&=\exp \left[4i\Theta(t)\left(J_{x}+1\right)\right]
            \nonumber \\ 
            &= \sum_{l=-\infty}^{\infty}\mathcal{J}_{l}\left[\frac{4\gamma^{x}_{1}}{N\Omega}(J_{x}+1)\right]e^{il\Omega t}
      .  \label{Exponential2}
\end{align}
%%%
To obtain Eqs. \eqref{Exponential1} and \eqref{Exponential2}, we have used the identity
%%%
\begin{equation}
      \label{BesselIdentity}
            \exp(iz\sin \Omega t)=\sum_{l=-\infty}^{\infty}\mathcal{J}_{l}(z)e^{il\Omega t}
      ,
\end{equation}
%%%
where $\mathcal{J}_{l}(z)$ is the $l$th-order Bessel function \cite{Abramowitz}.

To complete the transformation into the rotating frame, let us consider $\hat{U}_{z}(t)=\exp \left(-i\theta_{m}(t)J_{z}\right)$, where $\theta_{m}=\frac{m\Omega t}{2}$. The composition of the two rotations corresponds to transformation into the rotating frame of Eq. \eqref{UnitaryLMGRot}:
%%%
\begin{equation}
      \label{TraRotZ}
            \rvert \Psi_{m},t\rangle=\hat{U}^{\dagger}_{z}(t)\rvert \Psi_{I},t\rangle=\hat{U}^{\dagger}_{m}(t)\rvert \Psi,t\rangle
      .
\end{equation}
%%%
Finally we obtain the Hamiltonian in the rotating frame of Eq. \eqref{HamiltonianRotFrameSI}:
%%%
\begin{align}
      \label{RotatingHamiltonian}
            \hat{H}_{m}(t)&= \hat{U}^{\dagger}_{z}(t)\left(\hat{H}_{I}(t)-i \frac{\partial}{\partial t}\right)\hat{U}_{z}(t) 
            \nonumber \\ 
            &=\hat{U}^{\dagger}_{z}(t)\hat{H}_{I}(t)\hat{U}_{z}(t)-\frac{m\Omega }{2}J_{z}
      .
\end{align}
%%%
The operators defined in Eqs. \eqref{FirstOperator} and \eqref{SecondOperator} are given by the identity $\hat{\mathcal{O}}^{m}_{1,2}(t)=\hat{U}^{\dagger}_{z}(t)\hat{\mathcal{O}}_{1,2}(t)\hat{U}_{z}(t)$. Correspondingly, the operators Eqs.\eqref{RotJY} and \eqref{RotJX} read 
%%%
\begin{align}
            \hat{\Lambda}^{m}_{1}(t)&=-\hat{U}^{\dagger}_{z}(t)J_{y}\hat{U}_{z}(t)
            , \label{Rotat1} \\
            \hat{\Lambda}^{m}_{2}(t)&=\hat{U}^{\dagger}_{z}(t)J_{x}\hat{U}_{z}(t)
     .  \label{Rotat12}
\end{align}
%%%
%%%%%%%%%%%%%%%%%%%%

%%%%%
\end{document}